\documentstyle[sprocl]{article}
\def\be{\begin{equation}}
\def\ee{\end{equation}}
\def\bea{\begin{eqnarray}}
\def\eea{\end{eqnarray}}

\newcommand{\e}{{\rm e}}

\begin{document}
\title{STRING THEORY AND MATRIX MODELS
\footnote{
Talk presented at the Workshop "Frontiers 
in Quantum Field Theory" in honor of the 60th 
birthday of Prof. Keiji Kikkawa, Osaka, Japan, December 1995}}
\author{ T. YONEYA }
\address{Institute of Physics, University of Tokyo, 
Komaba, Meguro-ku, Tokyo 153, Japan}
\maketitle\abstracts{
It is generally accepted that the double-scaled 1D matrix model  
is equivalent to the $c=1$ string theory with 
tachyon condensation. There remain however 
puzzles that are to be clarified in order to 
utilize this connection for our quest towards 
possible non-perturbative formulation 
of string theory.  We discuss some of the issues that are 
related to the space-time interpretation of 
matrix models, in particular, the questions of leg poles, causality, 
and black hole background. Finally, 
a speculation  about  a possible  
connection of a deformed matrix model 
with the idea of Dirichret brane is presented.}
\section{String theory from matrix models}
There are many reasons for believing that in string theory 
non-perturbative effects play crucial roles 
at various stages in constructing reasonable 
solutions.  First of all, the space (moduli space) of   
the perturbative vacua of string theory is so rich that 
its structure should be formulated as a sort of generalized dynamical 
theory beyond perturbation theory. More importantly, 
general string perturbation 
series with respect to string coupling constant 
$g_{{\rm st}}$ are badly divergent (non Borel summable), 
in fact worse than those 
in renormalized perturbation series in local field theories. 
This is so in spite of an important fact that 
each term of the series is ultraviolet finite 
in contrast with local field theory, especially, 
that of quantized Einstein 
theory.  A phenomenon closely related with this 
is \cite{shenker} that 
a typical non-perturbative effect is expected 
to be of the form $\exp -1/g_{{\rm st}}$ 
instead of $\exp -1/g_{{\rm st}}^2$.  Both of these 
can be seen clearly in string theory interpretation 
of matrix models. Furthermore, 
we expect that, if non-perturbative effects 
are so crucial in string 
theory, the theory is not even of the theory of strings but 
is described  by more fundamental degrees of freedom.  
In this respect, matrix models are quite suggestive, since  
the original degrees of freedom  here are infinite dimensional 
matrix fields which are 
defined {\bf locally} in target space. 

%The way the matrix model 
%simulates string theory, however, appears quite mysterious, 
%and there remain several controversial issues on 
%its interpretation. 
%In  the following, I would like to explain them, try to 
%make some clarifying remarks and speculate future 
%possibilities. 

The basic reason why we can expect that matrix models 
may be interpreted  as string theory \cite{review} comes from the 
random surface interpretation.  The continuum limit 
of random surfaces is described by 2D Liuoville gravity 
in which the target space of the matter fields is just 
the base space of the matrix fields.  The Liouville degree of freedom 
can  in turn be regarded as an additional target space coordinate, 
provided appropriate background condensations 
of dilaton $\Phi$ and/or  
tachyon $T$  
string modes are assumed.  

The so-called linear dilaton background 
is the simplest exact vacuum solution for string 
theory with 2D flat space-time 
background. 
\be
\Phi= {2\over \sqrt{\alpha'}}x^1, \quad
T =0, \quad 
G_{\mu\nu}=\eta_{\mu\nu}
\ee
where $x^1$ is the spatial coordinate. 
\footnote{Our target space metric here is Minkowski. }
The string coupling constant is determined by the 
dilaton condensation as 
$ g_{{\rm st}} \propto \exp \Phi \sim \exp 2x^1
/\sqrt{\alpha'}$. This implies that the perturbation 
theory around this vacuum cannot 
be well defined, since the coupling grows indefinitely 
as $x^1 \rightarrow 
\infty$.  Fortunately, however, 
the linear dilaton vacuum can be deformed 
without violating 
conformal invariance by letting tachyon condense in the form 
\be
T= \mu \e^{2 x^1 / \sqrt{\alpha'}}.
\ee
The infinite repulsive wall 
formed by the tachyon condensation in the region 
 $x^1 \rightarrow \infty$ 
prevents strings going into the strong-coupling region and saves 
the above difficulty. In the opposite asymptotic region 
where $x^1 \rightarrow -\infty$, the string coupling is exponentially 
vanishing. Hence strings are appreciably interacting only in the wall 
region and overall coupling strength is  proportional 
to ${1 \over \mu}$, as can be easily seen by a scaling argument 
shifting the origin of the Liouville coordinate $x^1$. 
The asymptotic particle content of 2D string theory is 
therefore exactly 
described by the free string theory in the linear dilaton vacuum. 
Only particle mode which can be adopted as asymptotic 
states for scattering experiment is the so-called massless tachyon. 
Other possible physical modes called discrete states can be 
interpreted as global degrees of freedom whose condensation 
may further deform the structure of the vacuum solution itself. 
The existence of discrete states is intimately 
related with 
another characteristic feature of the 2D string 
theory.  Namely, the discrete states can be 
associated with an infinite 
number of conserved currents which form the 
$W_{\infty}$ algebra.  

Now let us turn to 1D matrix model.  The action for an Hermitian 
$N\times N$ 
matrix $M(t)$ is  
\be
S=\int dt {N\over 2}\,{\rm Tr}(\dot M(t)^2 + M(t)^2). 
\ee
The inverted harmonic oscillator potential  is treated 
with some finite-volume cutoff at $\lambda=-A$ and taking 
the limit $A\rightarrow \infty$ limit together with 
$N\rightarrow \infty$ 
afterwards, where $\lambda$ 
is the eigenvalue of the matrix $M$. 
If we assume that the physical states of the model are $U(N)$ singlet, 
the model is equivalent with a system of $N$ non-interacting fermions  
in the potential $V(\lambda) = -{1\over 2}\lambda^2$. 
Then the ground state is parametrized by the fermi energy $e_F$ 
measured downward from the top of the potential. 
In the limit $N \rightarrow \infty$, excited states  
are described as a collective mode of the 
surface excitation of the fermi sea. In the asymptotic region 
$\vert \lambda \vert \rightarrow \infty$, it is a free massless field. 
The effective field theory (collective field theory) for 
this mode shows that the overall strength of interaction is given by 
${1\over Ne_F}$. Furthermore, it turns out that the tree and one-loop 
contributions 
to the ground state 
energy has logarithmic dependence $\log e_F$ 
apart from the usual dependence $(Ne_F)^{2-2p}$ for genus $p$ 
contribution. Under the identification $\mu \sim Ne_F$, 
these properties are just the behavior 
expected from the presence 
of linear-dilaton and tachyon background and strongly suggest 
the equivalence of the model with 2D string theory \cite{pol}. 
In particular, the massless collective excitation 
is then identified with the massless tachyon of 
2D string. The infinite wall formed by the tachyon 
condensation for the latter case is replaced by the 
potential wall for the former  
with positive $\mu$. 

However the final and only legitimate handle for 
making more precise identification between the matrix model 
and string theory is the $S$-matrix. In the matrix model, 
the $S$-matrix element can be rigorously calculated 
to all orders with respect to the strength of interactions 
\cite{dijk}. 
On the other hand, the corresponding computation in 
string theory is very difficult, because of the 
presence of tachyon condensation. It requires 
precise determination of correlation functions 
for the quantum Liouville theory, which has been an  
open question for more than a decade.
One possible approach to this problem has been the 
method of {\it analytic continuation} with respect the 
number of the insertions of the tachyon condensation 
operator $\mu \e^{2 x^1 / \sqrt{\alpha'}}$.  
The result of such computation 
\cite{gli} 
indicates that the string-theory $S$-matrix 
element $A_{{\rm string}}
$ are proportional to the matrix-model 
$S$-matrix element $\overline{A}_{{\rm matrix}}$ 
apart from the energy($\omega$)-dependent 
phase factor $\ell(\omega)$  
for each external leg. Although they are pure phases, 
the leg factors, being energy dependent, cannot be discarded since 
they affect the space-time trajectories of 
the strings.   
\bea
&&A(\omega_1,\ldots, \omega_n \rightarrow 
\omega_{n+1},\ldots, \omega_{n+m})_{{\rm string}}
\nonumber
\\
&=&
\{ \prod_{i=1}^{n+m} \ell(\omega_i) \}
\overline{A}(\omega_1,\ldots, \omega_n \rightarrow 
\omega_{n+1},\ldots, \omega_{n+m})_{{\rm matrix}}
\eea
where we are considering the $n ({\rm in}) \rightarrow m ({\rm out})$ 
scattering and  
\be
\ell(\omega) = \mu^{-\sqrt{\alpha'} i\omega/2}
{\Gamma (i\sqrt{\alpha'} \omega)
\over 
\Gamma (-i\sqrt{\alpha'} \omega)}.
\ee
The matrix model $S$-matrix $\overline{A}_{{\rm matrix}}$ 
have a 
genus expansion, 
\be
{\overline A}_{{\rm matrix}}= {1
\over \mu^{m+n-2}}
(a_0(\sqrt{\alpha'}\omega) + {1 \over \mu^2} a_1(
\sqrt{\alpha'}\omega)
 + \cdots)
\ee
where the  genus$-p$ amplitude $a_p$ is a piecewise 
polynomial with respect to the energies and 
has no cut and/or pole singularities usually 
expected for the perturbative $S$-matrix elements.  
Higher genus contributions get additional powers of 
energies as $(\omega/ \mu)^{2p}$. 
Another important property is that $a_p$ always vanishes for 
vanishing energies for $n+m>2$, even if the kinematical 
energy factors coming from the relativistic normalization 
for the wave function are 
separated out. 
  
Thus the singularities of the  string $S$-matrix are 
only contained in the leg factor $\ell(\omega)$. 
This peculiar singularity structure of 2D string 
theory is a consequence of the very special kinematics 
of  asymptotic massless tachyons in the linear dilaton 
background.  
Although we have no complete proof for the correctness of 
the above $S$-matrix, there are further evidences 
which strengthen the result.  For instance, the 
above structure is consistent with the Ward-like 
identities associated with the $W_{\infty}$ 
currents. We refer the audience to Hamada's talk \cite{hamada} 
in this 
meeting about a derivation of the S-matrix using the 
$W_{\infty}$ Ward identities.

\section{Bulk versus wall scattering: 
Problem of causality?}

 A simple but important 
consequence of the above structure 
of the $S$-matrix is that the local-field limit, 
namely, the zero-slope limit ($\alpha' \rightarrow 0$) 
is trivial for 2D strings. The $S$-matrix element vanishes 
except for the 2-point amplitude which is one apart from a  
normalization factor. 
In contrast with the cases of critical strings, 
there is no nontrivial local-field limit in the 
systematic expansion with respect to $\alpha'$. This 
is natural if we remember that 
the effective string coupling  
$\exp 2x^1
/\sqrt{\alpha'}$ 
just becomes an infinite vertical wall at the origin $x^1=0$ and 
for $x^1 <0$ the interaction vanishes. In this sense, 
all nontrivial properties of 2D strings 
should be understood as a consequence of string extension 
in spite of the fact that there is no transverse 
extension of strings.  

Keeping this in mind, let us first try to 
interpret the singularities of the leg factor 
from the point of view of ordinary local field theory. 
We first notice that the positions $\omega=
{in \over \sqrt{\alpha'}} \, \, (n=1, 2, \ldots)$ of the poles 
in the leg factor coincide with those of the discrete 
states. In fact, the 
operator product expansion of the vertex operators 
leads to these poles 
as in the ordinary critical string theories, if we 
neglect the effect of tachyon condensation. 
These poles can also be interpreted as being due to the 
resonance between incident waves with the 
tachyon background which has a pure imaginary momentum 
$p=i2/\sqrt{\alpha'}$. For example, 
for n $\rightarrow$ 1 scattering with pure imaginary 
energies, the resonance condition with $t$ insertions 
of tachyon condensation is 
\be
\sqrt{\alpha'}\omega_{n+1}= i(n+t-1)= \sum_{i=1}^n \omega_i.  
\label{tachyonresonance}
\ee
Furthermore, in the tree approximation, 
the residue at these poles are precisely 
given by the string amplitudes for the linear dilaton 
background with $t$-insertions of tachyon condensation 
operators.  In particular, the amplitudes for $t=0$,  
sometimes called bulk amplitudes, 
are just the tree $S$-matrix elements for the 
linear-dilaton vacuum without tachyon background. 
Natsuume and Polchinski \cite{natsupol} interpreted this phenomenon 
from a view point of classical space-time physics as follows. 
Fourier-transforming with respect to $\omega_{n+1}$ (out-going) 
and taking the early time limit $x_{n+1}^0 \rightarrow -\infty$, 
we find its leading  behavior is given as
\[
A_f(x_{n+1}^0+x_{n+1}^1)
\]
\[
 \equiv \int_{-\infty}^{\infty} 
d\omega_{n+1}{\overline f(\omega_{n+1})}
A(\omega_{n+1})\e^{i\omega_{n+1}(x_{n+1}^0+x_{n+1}^1)}
\]
\be
\e^{(n-1)(x_{n+1}^0+x_{n+1}^1)}\times [
{\rm residue \, \, of} \, \, A(\omega_{n+1})\,\, 
{\rm at}\, \, \sqrt{\alpha'}\omega_{n+1}=i(n-1)].
\label{earlytimestring}
\ee
Namely, the leading early-time limit of the 
string tree amplitudes are proportional to 
the bulk amplitudes. This is natural 
since in the early-time limit, the string interaction 
is occurring only in the asymptotic region 
of space-time, {\it provided} that the incident wave packets 
are sufficiently localized. Thus the bulk amplitudes 
are exponentially small tails of the string amplitudes,  
whose main contributions are actually wall contributions. 

From the matrix-model point of view, the leg factor 
is a nonlocal field redefinition for the massless 
tachyon. 
Its early 
time $x^0\pm x^1 \rightarrow -\infty$ behavior is 
\be
\tilde \ell (x^0\pm x^1) \equiv \int_{-\infty}^{\infty} 
\omega \e^{i\omega(x^0\pm x^1)}
\ell (\omega) \sim \e^{(x^0\pm x^1)/\sqrt{\alpha'}}.
\label{earlytimematrix}
\ee
In the free-fermion picture, on the other hand, what is occurring  
is simply a potential scattering. 
A comparison of (\ref{earlytimematrix}) 
with (\ref{earlytimestring}) shows that the 
exponential tail is produced entirely 
by the field redefinition corresponding to 
the leg factor. Furthermore, to reproduce precisely the 
bulk amplitudes, it is crucial \cite{pol2} that 
the $W_{\infty}$ charges are conserved in the 
potential scattering. 
This then raises a subtle question about causality 
in the matrix model interpretation of string scattering 
when the incident wave is not small such that 
the height of the incident wave exceeds 
the top of the wall. Since the $W_{\infty}$ 
charges are not conserved in this case, 
we cannot have correct bulk scattering from the 
matrix model. On the other hand from the 
usual space-time view point, even if the amplitude is large, 
the interaction can be arbitrarily small 
in the asymptotic region, and the perturbative 
approximation for the buld amplitude 
should be valid provided the 
wave packet is sufficiently localized, since then  
the wave packet do not reach to the wall region. 
Does this indicate 
\cite{pol2} that the matrix-model 
interpretation of string theory necessarily 
violate causality?
%\vspace{0.1cm}
%\noindent
I would like to add three remarks related with 
this question, although they do not answer the question directly. 
\begin{enumerate}
\item The classical space-time picture explained above 
is only valid if we neglect the positivity of energy in quantum theory. 
I will argue that the bulk amplitudes cannot 
be the leading behavior even in the early time limit 
for the system with massless particle. 
\item We should take into account the effect of 
string extension, since, as I have emphasized before,  
all nontrivial behaviors of the string 
amplitudes should actually be consequences 
of string extension. 
\item Finally, there is a natural modification 
of the matrix model potentials such that 
the $W_{\infty}$ charges are conserved 
even for large amplitudes. It is plausible that 
the modified model describes a black hole background.  
\end{enumerate}
In the remainder of this section, 
the first point will be discussed. I will comment on the 
second and third points later. 

%\vspace{0.3cm}
It is well known that if only the $S$-matrix 
elements are given, we can only talk about, at best, the so-called 
{\it macro  causality}. Namely, causality is valid only within 
exponentially small errors, because of the impossibility of localizing 
 particle positions in relativistic quantum theory. 
For theories with massless particles, the situation is worse since 
we can only 
localize the wave packet with power behaving tails. 
Consider a wave packet of out-going asymptotic state
\be
\psi(x^0+x^1)=\int_0^{\infty} d\omega f(\omega)\e^{-i\omega (x^0+x^1)} 
\ee
where $f(\omega)$ is peaked around some value $\omega_0$. 
Here it is crucial to note that the  
range of integration for $\omega$ is the positive real axis, 
since negative energy is not allowed  
for the asymptotic particle states. 
Assuming that $f(\omega)$ is analytic near 
$\omega=0$ but vanishes rapidly for 
$\omega \rightarrow i\infty$, and has a pole at $\sqrt{\alpha'}\omega =
i$, 
\be
\psi(x^0+x^1)\sim -\int_{-\infty}^0 
d\omega f(\omega)\e^{-i\omega (x^0+x^1)}
+ 2\pi i {\rm Res}f(\omega)\Biggr\vert_{\sqrt{\alpha'}\omega=i}
\times  
\e^{x^0+x^1} 
\ee
But a general theorem says
\be
\int_{-\infty}^0 d\omega f(\omega)\e^{-i\omega (x^0+x^1)} 
\sim {\rm O}({1\over \vert x^0+x^1\vert^{\eta}})
\ee
as $x^0 \rightarrow -\infty$ for some $\eta\ge 1$ where 
$\eta$ is determined by the behavior of $f(\omega)$ 
near the origin $\omega =0$. For example, 
$\eta = 1$ for $f(\omega)\sim$ non-zero constant at $\omega =0$. 
Thus the leading term of $\tilde  
A(x^0+x^1 )$ in the early-time limit is suppressed only by 
a power behaving term. This shows that the 
bulk amplitudes can never be clearly separated 
even in the early-time limit in massless theories.  
In the space-time picture, we can easily imagine that 
such a power-behaved contribution 
comes from processes associated with the pair creation 
of antiparticles near the wall.  Remember that 
the apparent causality violation in quantum 
field theory is in general due to the existence 
\footnote
{See for example Feynman's lecture \cite{feynman} 
on why there must be antiparticles. 
}
 of antiparticles, 
corresponding to particles traveling backward in time. 
Causal propagator gives power behaving 
contribution outside light cone for massless fields.

\section{String extension and scattering phase shift}

The effect of string extension would make the issue 
of causality even more subtle.  
Unfortunately, however, we do not have 
appropriate space-time formulation of 2D string 
theory which properly takes into 
account the string extension. 
Here, I briefly describe a sample calculation 
\cite{jly} of  
scattering phase shift, to exhibit a dramatic role of 
string extension. 

Consider 2-point amplitude $A(\omega\rightarrow \omega)$ 
in the usual world sheet picture, 
\be
S = {1\over 4\pi\alpha'} \int d^2\xi 
\sqrt{{\hat g}} \biggl\{ \hat{g}^{ab}
\partial_a X_{\nu} \partial_b X^{\nu}  
-2\sqrt{\alpha'} \, \hat{R}^{(2)}x^1 
+ \mu \, 
\e^{2x^1/\sqrt{\alpha'} }
\biggr\}
\ee
with vertex operators 
\be
T_{\pm \omega} = e^{- i\omega x^0 (z , \bar{z} )} \, \e^{(-{2
\over \sqrt{\alpha'}} \pm i\omega )x^1(z , \bar{z} )}.
\ee

Simple perturbative 
expansion with respect to $\alpha'$ does not work, 
nor is with respect to $\mu$. 
Fortunately, however, 
in the limit of $\omega\rightarrow \infty$, 
the one-loop 
approximation gives a correct result. 
Perform standard semi-classical 
calculation by making a shift $x^1 
\rightarrow x^1_{{\rm classical}} + \tilde x^1$.
Classical (WKB) amplitude is given by
\be
\e^{iS_{{\rm classical}}},
\ee
\be
S_{{\rm classical}}= -\sqrt{\alpha'}\omega \ln(\mu/2) 
+ 2\sqrt{\alpha'}\omega(\ln(\sqrt{\alpha'}\omega)-1). 
\ee
This just corresponds to the center of mass motion of 
the string. 
One-loop contribution corresponding to the 
longitudinal fluctuation of the string for large $\omega$ is 
\be
e^{iS_{{\rm 1-loop}}},
\ee
\be
S_{{\rm 1-loop}} =  2\sqrt{\alpha'}\omega(\ln(\sqrt{\alpha'}\omega)-1).
\ee
The final result $\e^{i(S_{{\rm classical}}+S_{{\rm 1-loop}})}$ coincides 
with the high-energy limit 
of the two-point amplitude 
\be
\mu^{-i\sqrt{\alpha'}\omega}\Biggl({\Gamma(+i\sqrt{\alpha'}\omega)
\over \Gamma(-i\sqrt{\alpha'}\omega)}\Biggr)^2
\ee
which is obtained by 
{\it analytic} continuation in the number of insertions 
of tachyon condensation operators. 

This clearly shows the double-Gamma structure 
of the two-point amplitude coming from the leg factors
 and an essential role of 
string extension in producing the leg factors. 
Note that 
the center of mass motion is not 
sufficient to describe 2D string dynamics, 
even though there is no transverse modes.

%\section{Various unsolved problems and possibilities}

\section{Deformed matrix model, black-hole background and D-brane}

Let us next discuss a possible modification 
of the standard matrix model. 
In the ordinary $c=1$ matrix model, the $W_{\infty}$ charges 
are conserved only for small amplitudes. 
There is, however, almost unique modification \cite{aj} of the 
potential which remedies the situation. 
\be
V(M) = -{1\over 2}{\rm Tr}M^2 \rightarrow 
{1\over 2}{\rm Tr}(-M^2 + {m\over M^2}).
\ee
Scaling property \cite{jy} of the deformation term 
$m{\rm Tr}(1/M^2)$ is just consistent with the 
deformation corresponding to black-hole mass 
of the conformal-field theory approach. 
Namely, the string coupling scales 
as $g_{{\rm st}} \sim {1\over \sqrt{m}}$.  
In CFT approach, this comes from the difference that 
 the $x^1$ dependence of the black-hole 
mass operator is $\exp{4x^1 \over \sqrt{\alpha'}}$ 
in contrast to $\exp{2x^1 \over \sqrt{\alpha'}}$ of the tachyon 
condensation corresponding to   
$g_{{\rm st}} \sim {1\over \mu}$.

The deformed model leads to a curious prediction \cite{jy} 
that the odd-point scattering amplitude vanishes.  
%This in particular would imply that odd-point bulk amplitudes 
%are precisely canceled by other 
%contributions originating from pair creation near the 
%black hole region.  
Note that this is 
not inconceivable since the bulk amplitudes 
can never be clearly separated as argued above. 
We may interpret this phenomenon as a consequence of the 
compactification of time in the 
Euclidean black hole corresponding to the temperature 
$T_H={3\pi \sqrt{\alpha'}}$ 
\cite{dijkver} which leads 
the discrete Euclidean energies $\omega_n = {2n\over 3\sqrt{\alpha'}}
$. It is easy to 
check that the resonance condition (\ref{tachyonresonance}) 
for the insertion of the black hole mass cannot be satisfied for  
odd point amplitudes. Equivalently, in the Euclidean black hole 
case, allowed discrete states must have even Euclidean energies 
$\omega = {2n\over\sqrt{ \alpha'}}$. We therefore expect 
that the leg factor should only exhibit pole singularities 
at even energies.  Related with this is the fact that 
the $W_{\infty}$ algebra is also reduced to half of the 
standard model. The odd energy currents are excluded.       

Next I would like to speculate on a 
new possibility of  
interpreting the deformation. 
Recently, various forms of duality relations in string 
theory is attracting great interest. In particular, 
Polchinski 
\cite{pol3} pointed out that the soliton-type excitations 
of string theory may be treated dynamically by introducing 
open strings with Dirichret boundary condition signifying the 
coupling of strings to certain extended dynamical object 
called `Dirichret branes' (D-brane). If the dimensionality  
of the string coordinates subject to 
Dirichret condition is $p$ and the boundary 
condition for the remaining coordinates are 
Neumann, the dimension of the D-brane is $d-p-1$ 
in the d-dimensional target space-time. 

Let us now consider the deformation term in the 
form 
\be
{\rm Tr} {1\over M^2} = \int_0^{\infty} d\ell \, 
{\rm Tr}\e^{-\ell M^2}.
\ee
In the random-surface interpretation, 
the operator ${\rm Tr}\e^{-\ell M^2}$ creates a 
macroscopic hole, roughly, of length $\ell \sim 
\exp {2x^1\over \sqrt{\alpha'}}$ on the world sheet. 
This amounts to setting a Dirichret condition 
for the boundary of the hole with respect to 
the spatial coordinate $x^1$. 
The $\ell$-integration means that the 
position of the corresponding zero-dimensional D-brane must be 
integrated over with a special weight. 
If this interpretation is correct, the black hole 
horizon is somehow replaced by specially  weighted D-branes.
\footnote{
After the conference, a very interesting paper 
\cite{vafastro} 
appeared pointing out that the black hole 
entropy may be derived from the D-brane picture 
for the extremal black hole.} 
It is a challenge to establish this 
intriguing possibility in a more concrete way. 
To make progress along this line, we need more precise understanding 
on both the space-time 
picture of the matrix model and the general dynamics of D-branes,   
in particular, the origin 
of the leg factor from the view point of the 
matrix model and the dynamics of D-branes in the presence 
of nontrivial dilaton condensation \cite{li}.  
\footnote{
I would like to mention here a possible analogy 
with the similar phase factor appearing in the 
scattering amplitudes of QED. It might be worthwhile 
to pursue this analogy by trying to construct 
some interacting fermion theory such that the   
S-matrix for bosonized excitations is just given
 by the free-fermion S-matrix 
multiplied with the leg factors. The fermion might then be 
interpreted as the field describing string solitons. 
} 
%At present, its form is 
%determined only by making a comparison 
%with the results of conformal-field caluculations. 
%Because of this, we do not know the precise form 
%of the leg factor for the deformed model. 

%\vspace{0.5cm}
To conclude, I would like to 
emphasize that 
matrix models have still many facets to be pursued and learned  
in seeking for possible nonperturbative framework 
of string theory.   
We must try to achieve  
unification of matrix-model methods and other approaches. 
For example, we have to further develop our 
understanding on the relation of matrix model 
approach and the string field theories, and on 
supersymmetric and higher dimensional generalizations 
of the matrix-model method. 
It would also be an interesting challenge to formulate 
string dualities (both T and S dualities)
 within the matrix-models or perhaps some generalized 
framework. 

\section*{Acknowledgments}
It was a honor for me to participate in this interesting meeting 
to celebrate Prof. Kikkawa's 60th birthday.


\section*{References}
\begin{thebibliography}{99}
\bibitem{shenker}
S. Shenker, in  Proc. Random Surfaces and Quantum Gravity, 
eds. O. Alvarez et al. (Plenum Press, 1991) .
\bibitem{review}
For reviews,  see,  e. g., 
I. Klebanov, Proc. Trieste Spring School 1991, eds. J. Harvey et al. 
(World Scientific, Singapore, 1991); J. Polchinski,  Les Houshes lecture, 
hep-th/9411028 (1994). 
\bibitem{pol}
J. Polchinski, Nucl. Phys. B346(1990)253.
\bibitem{dijk}
R. Dijkgraaf, G. Moor and M. Plesser, Nucl. Phys. B425(1991)356. 
\bibitem{gli}
M. Goulian and M. Li, Phys. Rev. Lett. 66 (1991) 2051.
\bibitem{hamada}
K. Hamada, in this volume and references therein. 
\bibitem{natsupol}
M. Natsuume and J. Polchinski, Nucl. Phys. B424 (1994) 137.
\bibitem{pol2}
J.~Polchinski, Phys. Rev. Lett. 74 (1995) 63.
\bibitem{feynman}
R. P. Feynman. in "Elementary Particles and The Laws of Physics" by 
R. P. Feynman and S. Weinberg 
(Cambridge University Press, 1987).
\bibitem{jly}
A. Jevicki, M. Li and T. Yoneya, Nucl. Phys. B448(1995) 277. 
\bibitem{aj}
J. Avan and A. Jevicki, Comm. Math. Phys. 150(1992) 149. 
\bibitem{jy}
A. Jevicki and T. Yoneya, Nucl. Phys. B411(1994) 64. 
\bibitem{dijkver}
R. Dijkgraaf, E. Verlinde and R. Verlinde, Nucl. Phys. B371(1992) 269. 
\bibitem{pol3}
J. Polchinski, hep-th/9510017 (1995)  and references therein. 
\bibitem{vafastro}
C. Vafa and A. Strominger, hep-th/9601029 (1996).  
\bibitem{li}
M. Li, hep-th/9512042.  
\end{thebibliography}
\end{document}